\documentclass[aps,prl,twocolumn,groupedaddress,superscriptaddress,showpacs]{revtex4-1}

\usepackage{graphicx}
\usepackage{color}
\usepackage{amsmath}
\usepackage{amsfonts}
\usepackage{amssymb}
\usepackage{dcolumn}

\begin{document}

\title{Optimal synchronizability of bearings}

\author{N. A. M. Ara\'ujo} \email{nuno@ethz.ch}
\affiliation{Computational Physics for Engineering Materials, IfB, ETH
  Zurich, Wolfgang-Pauli-Strasse 27, CH-8093 Zurich, Switzerland}

\author{H. Seybold} 
\affiliation{Department of Earth, Atmospheric, and
  Planetary Sciences, MIT, Cambridge, MA 02139, USA}

\author{R. M. Baram} 
\affiliation{Center for Theoretical and
  Computational Physics, University of Lisbon, 1649-003 Lisboa,
  Portugal}

\author{H. J. Herrmann} 
\affiliation{Computational Physics for Engineering Materials, IfB, ETH
  Zurich, Wolfgang-Pauli-Strasse 27, CH-8093 Zurich, Switzerland}
\affiliation{Departamento de F\'isica, Universidade Federal do
  Cear\'a, 60451-970 Fortaleza, Cear\'a, Brazil}

\author{J. S. Andrade Jr.} 
\affiliation{Departamento de F\'isica,
  Universidade Federal do Cear\'a, 60451-970 Fortaleza, Cear\'a,
  Brazil}
\affiliation{Computational Physics for Engineering Materials, IfB, ETH
  Zurich, Wolfgang-Pauli-Strasse 27, CH-8093 Zurich, Switzerland}
 
\begin{abstract} 
Bearings are mechanical dissipative systems that, when perturbed, relax
toward a synchronized (bearing) state. Here we find that bearings can be
perceived as physical realizations of complex networks of oscillators
with asymmetrically weighted couplings.  Accordingly, these networks can
exhibit optimal synchronization properties through fine tuning of the
local interaction strength as a function of node degree [Motter, Zhou,
and Kurths, Phys. Rev. E {\bf 71}, 016116 (2005)]. We show that, in
analogy, the synchronizability of bearings can be maximized by
counterbalancing the number of contacts and the inertia of their
constituting rotor disks through the mass-radius relation, \mbox{$m\sim
r^{\alpha}$}, with an optimal exponent $\alpha=\alpha_{\times}$ which
converges to unity for a large number of rotors.  Under this condition,
and regardless of the presence of a long-tailed distribution of disk
radii composing the mechanical system, the average participation per
disk is maximized and the energy dissipation rate is homogeneously
distributed among elementary rotors.
\end{abstract}


\maketitle

A coherent synchronized motion can naturally emerge in a network of
oscillators when the coupling intensity exceeds the synchronization
threshold \cite{Pikovsky03,Acebron05,Osipov07,Boccaletti08}.
Synchronization is the mechanism responsible for numerous phenomena such
as, e.g., the vital contraction of cells producing the heartbeats, the
harmony in an orchestra, and the coherence of an audience clapping after
a performance \cite{Neda00,Winfree02,Strogatz03}. However, undesired
synchronization might also be responsible for neural diseases and
collapse of technical infrastructures and networks
\cite{Strogatz05,*Floyd93,*Glass01,*Louzada12}.  Therefore,
understanding how synchronization can be enhanced or mitigated is a
question of paramount importance. The properties of the transition to a
synchronized state are known to be a result of the interplay between the
dynamics of the oscillators and the complex topology of the system
\cite{Assenza11,GomezGardenes11}. Previous studies have shown that
synchronization can be enhanced on scale-free topologies by asymmetric
weighted couplings, in contrast to random graphs, where the most
efficient configuration corresponds to a uniform coupling strength
\cite{Motter05b,*Motter05}.  More precisely, by expressing the
interaction strength $s_i$ of site $i$ in terms of its degree $k_{i}$ as
$s_{i}\equiv k_{i}^{-\beta}$, where $\beta$ is a tunable parameter,
Motter {\it et al.} \cite{Motter05b,*Motter05} observed that the
properties of the coupling Laplacian matrix
\cite{Heagy95,*Pecora98} lead to optimal synchronization at $\beta=1$.
Under this condition of maximum synchronizability, the coupling strength
just counterbalances the number of connections, thus minimizing the
total cost associated with the network of couplings.
\begin{figure}
\includegraphics[width=0.9\columnwidth]{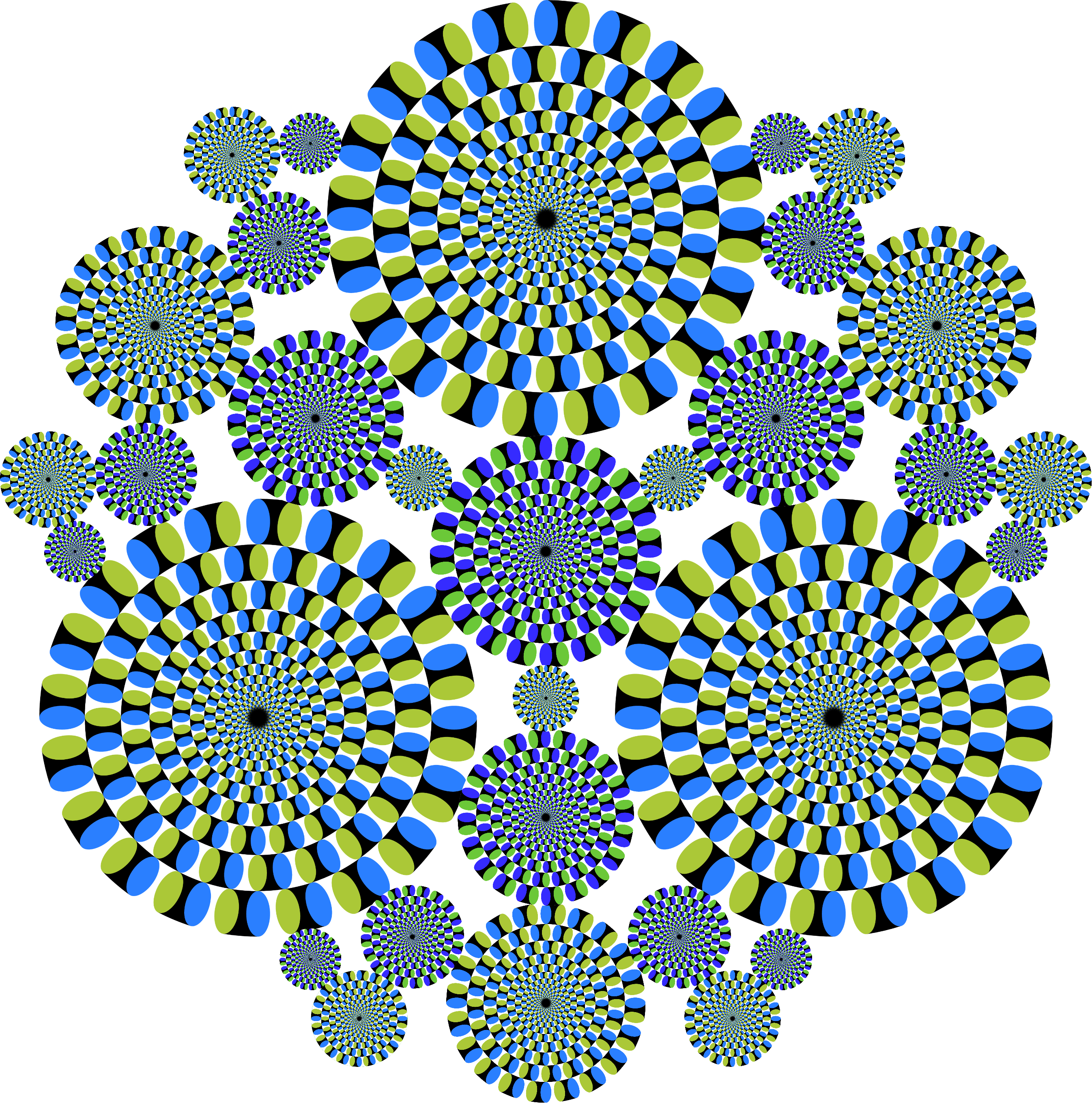}
\caption{Two-dimensional space-filling bearing configuration with $31$
rotor disks. In order to suppress gliding friction, disks rotate either
clockwise or anticlockwise, all with the same tangential velocity, and
loops of touching disks always have an even number of disks. This static
illusory-motion image is an adaptation from the peripheral drift
illusion ``Rotating Snakes'' shown in
Ref.~\cite{Kuriki08,*AkiyoshiSite}. Note that every two touching disks
always have opposite senses of rotation.  \label{fig::bearing}}
\end{figure}

Space-filling bearings have been previously considered to explain the
existence of seismic gaps \cite{Roux93}, which are those regions between
tectonic plates where no earthquake activity has been detected for a
large period of time \cite{Lomnitz82}. The idea is that the system
self-organizes into a ``bearing state'' in which the fragments rotate
without gliding friction. Systematic procedures have then been proposed
to generate model bearing structures of rotors with circular and
spherical shapes, in two and three dimensions, respectively, either
highly symmetric \cite{Herrmann90,Baram04} or random \cite{Baram05}. As
depicted in Fig.~\ref{fig::bearing}, hierarchical space-filling packings
emerge from these models in 2D, where the interstices among large disks
are sequentially filled by smaller ones. Gliding friction is suppressed
by ensuring that loops of touching disks have an even number of
constituents. In this way, clockwise turning disks only touch
counterclockwise rotating ones and vice-versa. At steady state, the
tangential velocity is the same for all contacts.
\begin{figure*}[t]
\begin{center}
\includegraphics[width=\textwidth]{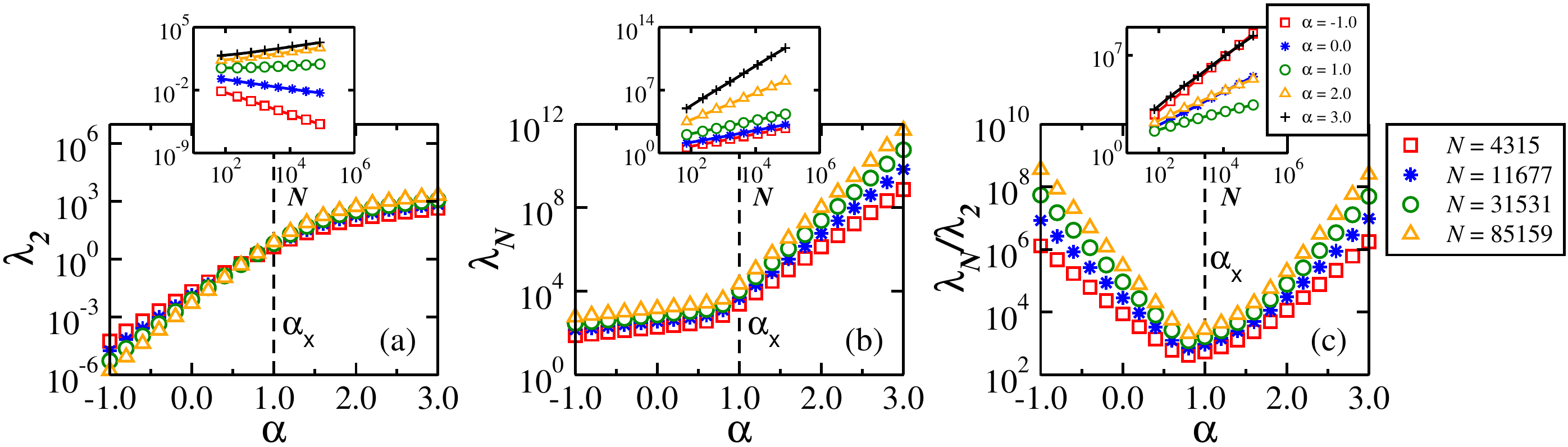}
\caption{Dependence on the exponent $\alpha$ of the lowest (non-zero)
$\lambda_2$ (a) and the largest $\lambda_N$ (b) eingenvalues of the
matrix $\mathbf{B}$. The same is shown in (c), but for the ratio
$\lambda_N/\lambda_2$. The insets show the system size dependence for
different values of $\alpha$. Both $\lambda_2$ and $\lambda_N$ increase
with $\alpha$ exhibiting a change in the behavior at the crossover value
$\alpha_{\times}$ (dashed lines): the faster increase of $\lambda_2$ is
for $\alpha<\alpha_{\times}$, while $\lambda_N$ grows faster for
$\alpha>\alpha_{\times}$. As a consequence, a minimum is observed for
the ratio $\lambda_N/\lambda_2$ at the crossover value
$\alpha_{\times}$.  \label{fig::eigenvalues.2d}}
\end{center}
\end{figure*}

Under a different framework, hierarchically filled structures, where
smaller elements (e.g, disks and spheres) are snugged into the
interstices of larger ones, have been directly associated to scale-free
networks \cite{Andrade05,Varrato11,Reis12,Zhang09,*Kaplan09,*Halu12}.
The Apollonian packing of circles, for example, inspired the
introduction of the so-called Apollonian network \cite{Andrade05},
where the sites correspond to the centers of the circles, and the edges
are drawn to connect the centers (sites) of pairs of touching circles.
Bearings can also
be directly associated to complex networks, whose sites are given by the
positions $\vec R_{i}$ of the centers of the disks. These spatial
networks fulfill, for each loop of $n$ disks, the condition
$\sum^{n-1}_{i=1}\left(\vec R_{i+1}-\vec R_{i}\right)=\vec R_{n}-\vec
R_{1}$, and are scale free, if the original bearing is space-filling.

One can readily identify the bearing state as a typical
synchronized state. It is thus legitimate to convey that space-filling
bearings rotating in steady-state (i.e., when all rotors possess equal
tangential velocities) are in fact physical realizations of synchronized
complex networks. Once this conceptual parallelism is ascertained, one
can go even further and ask, in the spirit of the asymmetric coupling
approach introduced in Ref.~\cite{Motter05b,*Motter05}, whether or not
such a synchronized state can be optimized through some constitutive
physical property of the bearings. In what follows we show that the
synchronized state of two-dimensional space-filling bearings can indeed
be substantially enhanced by adequately adjusting the inertial
contribution of individual rotors to the global motion of the system.
\begin{figure}[b]
\begin{center}
\includegraphics[width=0.8\columnwidth]{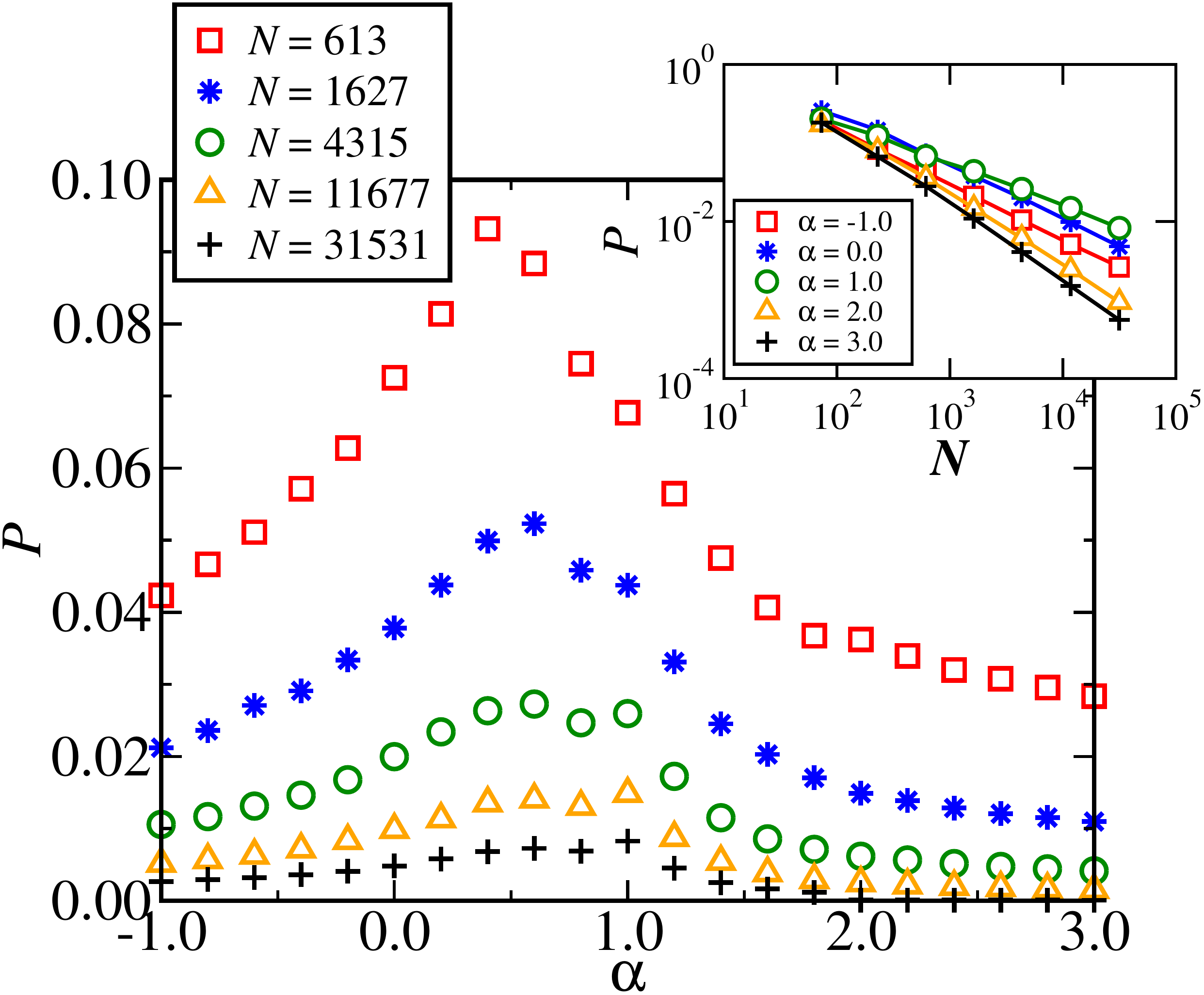}
\caption{$\alpha$-dependence of the average participation $P$
for bearings with different number of disks. An enhanced participation
per disk is observed for an intermediate range of $\alpha$. Results
suggest a discontinuous change in the position of the maximum as
discussed in the text. In the inset, the size dependence is included for
different values of $\alpha$. The participation per disk decreases with
the number of disks with the smallest slope for $\alpha=1.0$.
\label{fig::participation.2d}}
\end{center}
\end{figure}
\begin{figure}[b]
  \includegraphics[width=0.8\columnwidth]{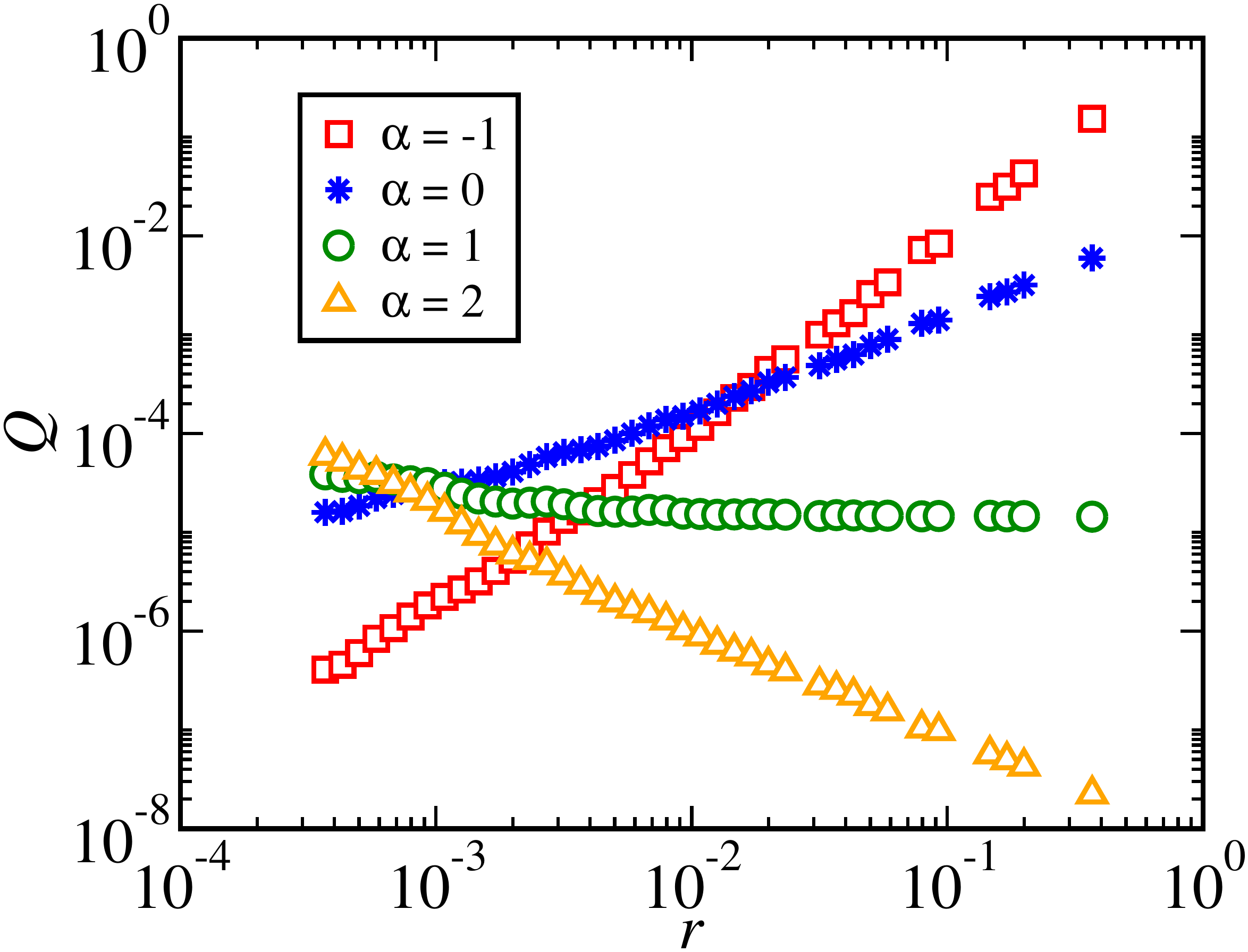}
\caption{Radius dependence of the relative dissipation rate $Q$ of each
disk for different values of $\alpha$, obtained for a bearing of
$85159$ disks. While for $\alpha<1.0$, $Q$ is an increasing function of
the disk radius $r$, for $\alpha>1.0$ larger disks dissipate less energy
as the number of contacts does not compensate the increase in disk
inertia.  \label{fig::dissipation} }
\end{figure}

Consider a bearing of $N$ rotors. The equation of motion for the angular
velocity $\vec{\omega}_i$ of rotor $i$ can be written as,
\begin{equation} \label{eq:motion}
 I_i \dot{\vec{\omega}}_i=\sum_j \vec T_j=\sum_j r_i\vec  r_{ij}\times
\vec F_{ji} \ \ , 
\end{equation}
where $I_i$ and $r_i$ are the rotational inertia and radius of rotor
$i$, the sum is over all rotors in contact with $i$, $\vec F_{ji}$ is
the force of rotor $j$ on the surface of $i$, and $\vec r_{ij}$ is the
unit vector pointing to the contact with $j$ in the reference frame of
disk $i$. Taking the force $\vec F_{ji}$ as a dissipative force
proportional to the relative velocity at the contact point, we have,
\begin{equation} \label{eq:force} 
\vec F_{ji}=\sigma (\vec v_j-\vec v_i)=-\sigma(\vec \omega_j \times
r_j\vec r_{ij} + \vec \omega_i \times r_i\vec  r_{ij} ) \ \ , 
\end{equation}
where $\sigma$ is the coupling between rotors and we used the identity
$\vec r_{ji}=-\vec r_{ij}$. In $2D$ bearings, since rotors are disks
with fixed position (see Fig.~\ref{fig::bearing}), the net translational
force is zero and the angular velocity can be described by a scalar.
Equation~(\ref{eq:motion}) then simplifies to,
\begin{equation} \label{eq::motion.simplified}
 I_i \dot{\omega}_i=-\sigma\sum_j A_{ij}\left[\omega_i r_i^2+ \omega_j
r_i r_j\right] \ \ , 
\end{equation}
where $A_{ij}$ are the elements of the connectivity matrix, defined in
such a way that $A_{ij}=1$ if two disks $i$ and $j$ are different and
mutually touching, and $A_{ij}=0$ otherwise.

At this point, we introduce the following constitutive relation between
the mass of disk $i$ and its radius,
\begin{equation}
 m_i=2 a r_i^{\alpha} \ \ ,
\end{equation}
 such that the rotational inertia becomes $I_i= a r_i^{\alpha+2}$, where
$a=1$ in consistent units, for convenience.  It then follows that, \mbox{$\dot{\omega}_i=
-\sigma\sum_j T_{ij}\omega_j$}, where $T_{ij}=n_i
r_i^{-\alpha}\delta_{ij}+r_i^{-1-\alpha}r_jA_{ij}$, $n_i$ is the number
of disks touching $i$, and $\delta_{ij}$ is the Kronecker delta. In
matrix form this can be written as,
\begin{equation} \label{eq::matrix_2d}
\dot{\vec{\omega}}=-\sigma\mathbf{T}\vec{\omega} \ \ ,
\end{equation}
where $\vec{\omega}$ is the $N$-dimensional vector of the angular
velocities and $\mathbf{T}$ is the interaction matrix. A bearing is a
dissipative system which, as already mentioned, converges to a
steady-state, namely the bearing state, where the tangential velocities
of all rotors become equal, i.e., $v_1=v_2=\cdots=v_N\equiv s$, such
that $\dot{s}(t)=0$. Through the relation between tangential and angular
velocities,  $\vec{\omega}=\mathbf{R}^{-1}\vec{v}$, where $\mathbf{R}$
is a diagonal matrix with $R_{ii}=c_ir_i$, with $c_i=\pm1$ depending on
the sense of rotation of the disk \cite{Baram04}, the equivalent to
Eq.~(\ref{eq::matrix_2d}) for the $N$-dimensional vector of absolute
tangential velocities $\vec{v}$ can be readily obtained, being
\mbox{$\dot{\vec{v}}=-\sigma\mathbf{B}\vec{v}$}. The coupling matrix
$\mathbf{B}$ can be written as,
$\mathbf{B}=\mathbf{R}\mathbf{T}\mathbf{R}^{-1}$, i.e.,
$B_{ij}=r_i^{-\alpha}(n_i\delta_{ij}-A_{ij})$, where we make use of
$c_i/c_j=-1$ for all pairs of touching disks. In this work we focus on
the relaxation after small perturbations $\vec{\xi}$ to the bearing
state, namely $v_i=s+\xi_i$, which leads to the following vectorial
variational equation:
\begin{equation} \label{eq::var_eq_2d}
\dot{\vec{\xi}}=-\sigma\mathbf{B}\vec{\xi} \ \ .
\end{equation}
This system of coupled linear differential equations can be written in
the space of eigenvectors of $\mathbf{B}$, $\vec{x}_k$ (with eigenvalue
$\lambda_k$), such that $\vec{x}_1$ ($\lambda_1=0$) refers to
perturbations along the stable manifold of the bearing state. All other
eigenvectors are transverse to $\vec{x}_1$ \cite{Pecora98}. Due to the
linear nature of Eq.~(\ref{eq::var_eq_2d}), the Lyapunov exponents
correspond to $-\sigma\lambda_k$. Since all eigenvalues are non-negative
in this problem (the matrix $\mathbf{B}$ can be symmetrized), the
stability of the bearing state is guaranteed \cite{Jordan07}. In analogy
to the work of Motter \textit{et al.} \cite{Motter05b,*Motter05}, the
factors $r_i^{-\alpha}$ present in the elements of $\mathbf{B}$
correspond to the weights of the pairwise interactions. As we show next,
in the complex network associated with the topology of the bearings, the
number of contacts is an increasing function of the radius and so it
becomes possible to enhance synchronization by tuning the coupling
weights in such a way as to balance the number of contacts.

Perturbations along the bearing states ($\dot{\vec{\xi}}=0$) lead the
system from one bearing state to another with a different tangential
velocity $s$, i.e., all tangential velocities change by the same amount,
regardless of the value of $\alpha$ ($\xi_i\equiv\xi$).  These
perturbations are related to the eigenvalue $\lambda_1=0$.  Hereafter,
we focus on perturbations which are transverse to the bearing states,
i.e., perturbations after which the system always relaxes back to the
original bearing state. Specifically, the objective is to investigate
the dependence on the inertial parameter $\alpha$ of the smallest
(non-zero), $\lambda_{2}$, and largest, $\lambda_{N}$, eigenvalues of
the system described by the matrix $\mathbf{B}$. These eigenvalues
correspond to the slowest and fastest relaxation modes, respectively.
Generally speaking, for perturbations out of the bearing manifold
($\lambda\neq0$), the larger the eigenvalues the faster the relaxation
toward the stable state. Here two features come into play: the mass
(inertia) distribution of the disks and the number of contacts. While
the former depends explicitly on $\alpha$, the latter is an increasing
function of the rotor radius. Numerical results for a 
bearing (of type $n=m=0$ of the first family for loops of size $4$
\cite{Herrmann90}) with $4511821$ disks reveal that the average number
of contacts scales with the disk radius as $r^\gamma$, with
$\gamma=0.94\pm0.04$, which should approach unity in the limit of space
filling systems.

As depicted in Figs.~\ref{fig::eigenvalues.2d}(a) and (b), $\lambda_{2}$
and $\lambda_{N}$ generally increase with $\alpha$, although changes in
the behavior of both eigenvalues can be observed at a crossover
value $\alpha_{\times}\approx 1$.  While $\lambda_{N}$ increases
faster for $\alpha>\alpha_{\times}$, the increase of $\lambda_{2}$
becomes attenuated in the same range of $\alpha$ values.  The insets of
Figs.~\ref{fig::eigenvalues.2d}(a) and (b) show that, for
$\alpha>\alpha_{\times}$, $\lambda_{2}$ and $\lambda_{N}$ are increasing
functions of the number of disks since the ones with higher inertia also
possess more contacts to shed any perturbation. The same description
applies for all eigenvalues.  For $\alpha<\alpha_{\times}$, the inertia
assigned to each disk does not always compensate its number of contacts.
For example, the fact that $\lambda_{2}$ decreases with $N$ for
$\alpha=-1$ reflects this imbalance. More general conclusions can be
drawn from the eigenratio between the largest and smallest (non-zero)
eigenvalues of the coupling matrix, $\lambda_{N}/\lambda_{2}$
\cite{Pecora98,Barahona02}.  This ratio solely depends on geometrical
features (radii, network topology, and disk mass) and not on the initial
conditions for the velocities. The lower the ratio, the higher the
synchronizability \cite{Pecora98,Barahona02}.  As shown in
Fig.~\ref{fig::eigenvalues.2d}(c), bearings consisting of rotors with
masses that follow the relation, $m\sim r^{\alpha_{\times}}$, have a
minimum eigenratio value, in analogy to the optimal synchronization
coupling found for scale-free networks \cite{Motter05b,*Motter05}.
Interestingly, the larger the value of $N$, the more sensitive is the
system to variations of $\alpha$.  Moreover, the results in the inset
indicate that, regardless of the value of $\alpha$, the eigenratio
increases monotonically with the number of disks. Nevertheless, since
this increase is less pronounced for $\alpha=1$ than for any other value
of $\alpha$, the relative depth of the minimum augments with system
size.  As a consequence, large systems display an enhanced relative
synchronizability at $\alpha=\alpha_{\times}$. In the \textit{Supplemental
Information} \cite{SM} we show the convergence of $\alpha_{\times}$ towards unity in
the thermodynamic limit.

To shed light on the dependence on $\alpha$ of the contribution of disks
to the eigenmodes, we define the average participation
\cite{Kramer93,*Bell70,*Morais11}
as,
\begin{equation} \label{eq::average.participation}
P=\frac{1}{N}\sum_{\vec{x}_k}\frac{\left[\sum_{j=1}
\vec{x}_{k}(j)^2\right]^2}{\sum_{j=1} \vec{x}_{k}(j)^4} \ \ ,
\end{equation}
where the outer sum is over all eigenvectors $\vec{x}_k$ and the inner
sums are over the components of the eigenvector. As shown in
Fig.~\ref{fig::participation.2d}, there is an intermediate range of
$\alpha$ values, for a given system size $N$, where an enhanced average
participation per disk can be clearly observed. However, our results
suggest that the value of $\alpha$ for which the participation becomes
maximum increases in a discontinuous fashion from $\alpha\approx 0.5$ to
$1.0$, as the system size increases from $N=613$ to $31531$.  The inset
of Fig.~\ref{fig::participation.2d} shows that, notwithstanding the
value of $\alpha$, the average participation per disk decreases with the
number of disks $N$, but with a slope that is milder for $\alpha=1.0$
than for any other case. This behavior expresses the more homogeneous
contribution of rotors to the system dynamics. In this range of $\alpha$
values, the number of contacts of each disk is approximately compensated
by its inertia. Accordingly, we expect $\alpha_{\times}$ to converge
toward $\gamma\approx1.0$ as the number of disks increases.  Under these
conditions, the impact of changes in the tangential velocity of a disk
becomes relatively less dependent on its radius.
 
\begin{figure}[t]
  \includegraphics[width=\columnwidth]{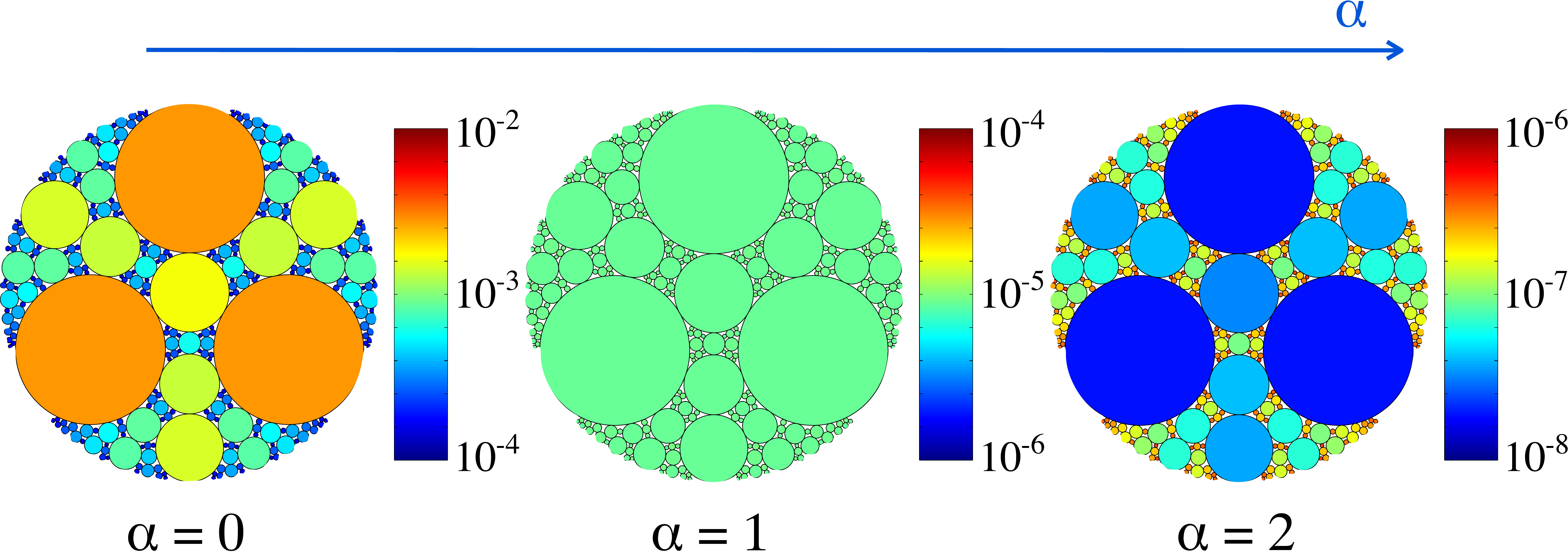}
\caption{Snapshots of bearings with $613$ disks and different values of
$\alpha$. The color stands for the average dissipation rate $Q$ in a
logarithmic scale. While for $\alpha<1$ the dissipation rate is an
increasing function of the radius, for $\alpha>1$ is a decreasing
function. For $\alpha=1$ the dissipation rate is of the same order of
magnitude to every disk.  \label{snap.dissipation}}
\end{figure}

It is also interesting to calculate the rate at which energy is
dissipated. For rotor $i$, we have \mbox{$W_i=I_iv_i\dot{v}_i/r_i^2$}.
Now, given an eigenmode $k$ with eigenvalue $\lambda_k$, the velocity of
disk $i$ can be expressed by the $i$-th component of $\vec{x}_k(i)$
scaled by a constant $c$, i.e., $v_i=c\vec{x}_k(i)$. Thus, one obtains
the relation $\dot{v}_i=-c\lambda_k\sigma \vec{x}_k(i)$. Based on this
result, we can decompose the dissipation rate of disk $i$ in its
different modes, $W_{ki}=-r_i^\alpha c^2\lambda_k\sigma \vec{x}_k(i)^2$.
It is then possible to quantify how dissipation is distributed among
disks, by defining the relative dissipation of disk $i$ in the eigenmode
$k$ as, $Q_i^k=W_{ki}/\sum_j W_{kj}$, which leads to, 
\begin{equation} \label{eq::relative.dissipation}
 Q_i^k=\frac{r_i^\alpha \vec{x}_{k}(i)^2}{\sum_j r_j^\alpha
\vec{x}_{k}(j)^2} \ \ .  
\end{equation}
Figure~\ref{fig::dissipation} shows, for different values of $\alpha$,
the dependence on the radius $r$ of the relative dissipation rate of a
rotor $i$ averaged over all eigenmodes, $Q_i=1/M\sum_kQ_i^k$, where $M$
is the total number of eigenmodes. For $\alpha<1.0$, the dissipation is
an increasing function of $r$, while, for $\alpha>1.0$, larger disks
dissipate less energy than smaller ones. By tuning $\alpha=1.0$, the
dissipation rate becomes more uniformly distributed among all disks,
being approximately invariant on the rotor size.
Figure~\ref{snap.dissipation} consists of snapshots for three different
values of $\alpha$, showing how the average dissipation rate is
typically distributed among disks in a bearing with $613$ disks.

In summary, we have shown that bearings are physical realizations of
complex networks of oscillators. When the bearings consist of rotors
of different sizes, the coupling between oscillators is asymmetric, as
the effect of a pairwise interaction on the rotors motion depends on
their inertia (typically different for each one).  Once this parallelism
is established, it is possible to evaluate the stability of these
mechanical systems applying concepts from dynamic systems theory. In
particular, our results for two-dimensional space-filling bearings,
characterized by a scale-free distribution of rotor (disk) contacts,
indicate that their synchronizability can be duly maximized.  This is
achieved by counterbalancing the number of contacts of the disks with
their inertia through the mass-radius relation,
$m\sim r^{\alpha_{\times}}$, where $\alpha_{\times}$ is the optimal
exponent which we expect to asymptotically converge to unity as the
number of rotors increases. Under this condition, in spite of the
power-law distribution of radii, the average participation per disk has
a maximum and the energy dissipation rate is homogeneously distributed
among disks. 

\begin{acknowledgments}
We thank Heitor Credidio for help in the preparation of
Figure~\ref{fig::bearing}.  We acknowledge financial support from the
European Research Council (ERC) Advanced Grant 319968-FlowCCS, the
Brazilian Agencies CNPq, CAPES, FUNCAP and FINEP, the FUNCAP/ CNPq
Pronex grant, and the National Institute of Science and Technology for
Complex Systems in Brazil. 
\end{acknowledgments}

\bibliography{bearings}


\section*{Supplemental Information}

\begin{center}
  \includegraphics[width=\columnwidth]{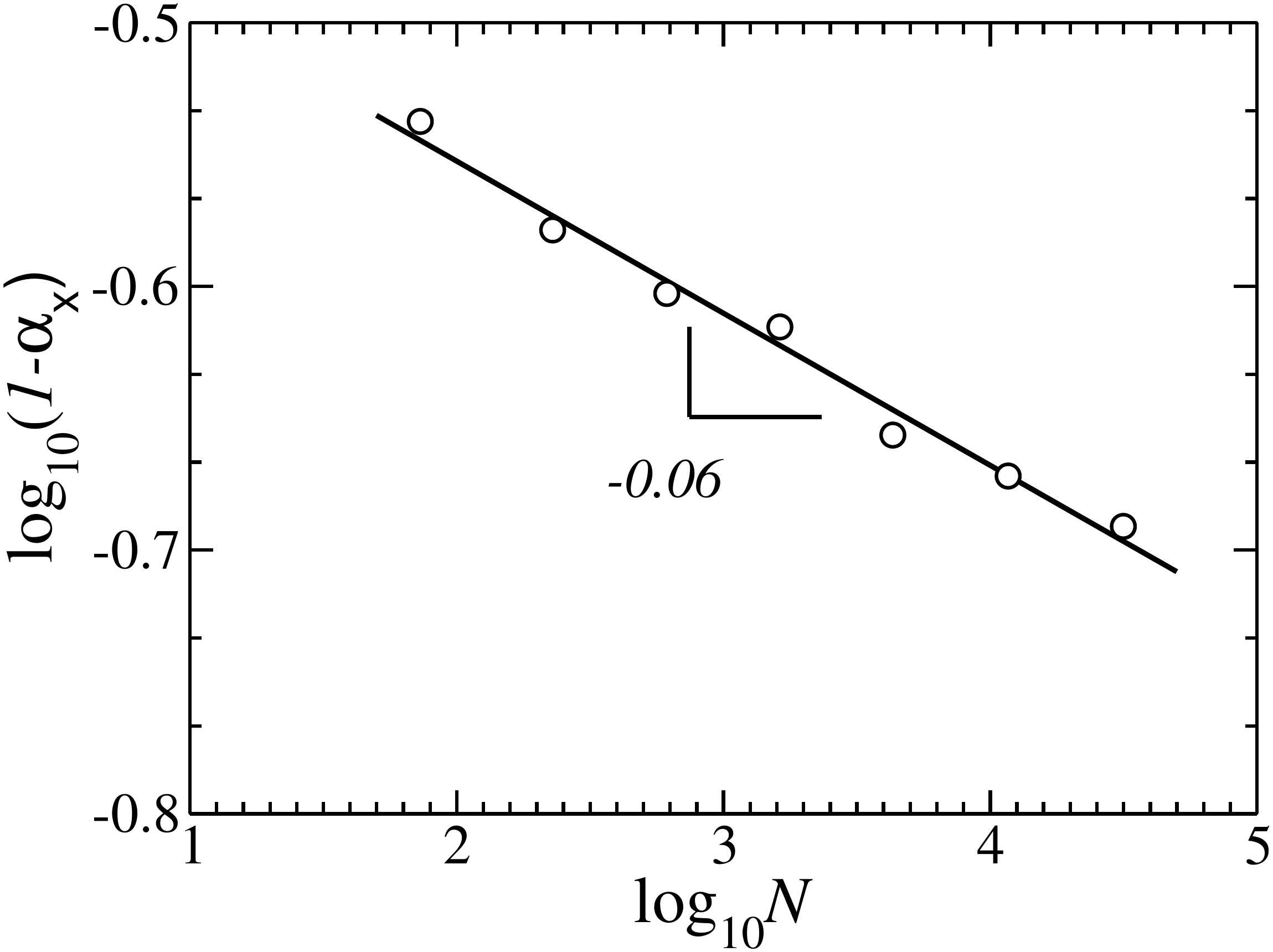} 
\end{center}
  Fig.~S1: Scaling of $1-\alpha_{\times}$ with the number of disks in the bearing, $N$. A power-law is observed
supporting the prediction of $\alpha_{\times}=1$ in the thermodynamic
limit.

\end{document}